\documentclass[%
 prereprint,
 amsmath,amssymb,
 aps,
]{revtex4-2}

\usepackage{graphicx}
\usepackage{dcolumn}
\usepackage{bm}

\usepackage{mhchem}
\usepackage {booktabs}
\usepackage {comment}

\begin{document}

\preprint{APS/123-QED}

\title{Simple measures to capture the robustness and the plasticity of soil microbial communities}

\author{Takashi Shimada}
\affiliation{Department of Systems Innovation, School of Engineering, The University of Tokyo}
\affiliation{Mathematics and Informatics Center, The University of Tokyo}
\author{Kazumori Mise}
\affiliation{Graduate School of Agricultural and Life Sciences, The University of Tokyo}
\affiliation{Graduate School of Science, The University of Tokyo}
\affiliation{Bioproduction Research Institute, National Institute of Advanced Industrial Science and Technology}
\author{Kai Morino}
\affiliation{Interdisciplinary Graduate School of Engineering Sciences, Kyushu University}
\author{Shigeto Otsuka}
\affiliation{Graduate School of Agricultural and Life Sciences, The University of Tokyo}
\affiliation{Collaborative Research Institute for Innovative Microbiology, The University of Tokyo}



\begin{abstract}
Soil microbial communities are known to be robust against perturbations such as nutrition inputs, which appears as an obstacle for the soil improvement.
On the other hand, its adaptable aspect has been also reported.
Here we propose simple measures for these seemingly contradicting features of soil microbial communities, robustness and plasticity, based on the distribution of the populations.
The first measure is the similarity in the population balance, i.e. the shape of the distribution function, which is found to show resilience against the nutrition inputs.
The other is the similarity in the composition of the species measured by the rank order of the population, which shows an adaptable response during the population balance is recovering.
These results clearly show that the soil microbial system is robust (or, homeostatic) in its population balance, while the composition of the species is rather plastic and adaptable.
\end{abstract}

\date{\today}

\maketitle

 
Understanding of the stability of ecosystems has been a central question in ecology, often with emphasis on local stability, persistence, permanence, resilience, etc., and its relation to its diversity~\cite{MacArthur1955, Gardner1970, May1972, Tregonning1979, Pimm1979, Pimm1984, Taylor1988A, McCann2000, Kondoh2003, McKane2005, Ings2009, MougiKondoh2012, FEMSRev2013}.
It has also been inspiring broader field to consider the robustness of complex systems which consist of many components interacting each other more in general~\cite{BakSneppen1993, Sole1996, Tokita1999, AlbertJeongBarabasi2000, Jain2002, Christensen2002, BuldyrevStanleyHavlin2010, Shimada2014SREP, Ogushi2017}.
However, empirical tests of various theoretical predictions on animal-plant ecosystems have been limited within natural experiment, i.e. comparing the systems under different environment and with different history. This is mainly due to the large scales in space and time scale, for conducting controlled experiments.

Microbial ecosystems are one of the best systems to resolve such situation, because they are easy to manipulate and their development can be observed in a shorter time scale.
Bacteria in soil are also known to be extremely abundant and diverse, with estimates generally ranging from $10^6$ to $10^{10}$ cells~\cite{trevors2010one} and $6,400-38,000$ taxa \cite{curtis2002estimating} per gram of soil. These bacteria play vital roles in biogeochemical cycling, plant growth, and the maintenance of terrestrial ecosystems. Reflecting the importance of soil functions for terrestrial ecosystems, there has been considerable effort invested in understanding the response of soil ecosystems to disturbance~\cite{griffiths2013insights}. 
The possibility of such approach was brilliantly illustrated in a pioneering work on the microbial communities of natural soil and natural water~\cite{ArtificialEcosystemSelection2000}.
In that study, it was shown that the macroscopic characteristics of the micro-ecosystems such as those dry weight and pH can be modulated by an artificial natural selection process.
And the difference was inherited to the following systems even after quitting the natural selection.
This shows that the microbial ecosystem have both plasticity and robustness or stability in macroscopic sense, while the underlying microscopic mechanism remained largely unknown.

Advance of DNA sequencing technology has enabled microbiologists to obtain detailed information on dozens of microbial communities at one time~\cite{Hamady2008,Caporaso2012}.
This has paved a way to elaborate time-series investigation of microbial ecosystem successions~\cite{Handley2015,Ridenhour2017}.
While many have elucidated the succession of microbial communities in a descriptive manner, for example by cataloguing a plethora of microbial clades (taxonomic groups) that increase or decrease during the ecosystem development~\cite{Samad2017,Jurburg2017}, others summarize the dynamics of the systems using statistical features such as
co-occurrence network indicators or ARIMA modeling~\cite{Faust2015,Ridenhour2017}.
Other characteristics such as the shape of species abundance distribution~\cite{Shoemaker2017} and the network structures of interactions~\cite{MarinoSchloss2014} have been also the target of investigation.
However, the interpretation of these features is by no means straightforward.

In this study, we propose simple and comprehensible measures to characterize the state of microbial communities, based on the balance and the composition of populations of the operational taxonomic units (OTUs).
These two measures are independent each other in the sense that there always exists a direction to modulate the community so that one measure changes while the other is kept constant.
It is shown that these measures well capture the conserved aspect (relating to the stability, robustness, etc.) and the adaptive aspect (plasticity) of the microbial community.

\section*{Results}
\subsection*{System and Preparation}
As a model system of community dynamics in response to perturbation, we here use time-series data of soil microbial communities receiving nutrient input~\cite{Mise2020}. This dataset consists of soil microbial community structures (populations of OTUs) at five time points, i.e. $0$, $3$, $10$, $17$ and $24$ days after initial preparation. As shown in TABLE \ref{table_RSA}, the nutrient input condition consists of the combination of nitrogen source: ammonium chloride (\ce{NH4Cl}, denoted by A in the labeling of the conditions) or urea (U), carbon source: glucose (G) or cellobiose (C). We also have two types of input schedule: either once in the beginning of the observation (day $0$) or addition of one quarter of the amount four times every week (days $0, 7, 14$, and $24$), denoted by $1$ and $4$ in the labeling of the condition, respectively. Negative controls without nutrient input were also prepared (``Control'').

This nutrient amendment schemes were designed to cover various intensities and types of perturbations.
Urea and cellobiose represent more slowly acting nutrition sources than ammonium and glucose, since the former two require enzymatic degradation before being catabolized by microbes.
Regarding the input schedules, the one-time and the four-time schemes represent large abrupt perturbation and small but continuous perturbation, respectively.

Since we have three samples for each condition, a soil sample $\alpha$ is specified by the combination of all these parameters as:
$\alpha = (t, d, s)$, where $n \in \{ {\rm Control, AG1,AG4, AC1, AC4, UG1, UG4, UC1, UC4} \}$,  $d \in \{ 0, 3, 10, 17, 24 \}$, and $s \in \{ 1, 2, 3 \}$ represent the treatment condition, observation day, and the sample number for each condition, respectively. Because the data $d=0$ is only for Control, the total number of the soil samples is $(1 \times 5 \times 3) + (8 \times 4 \times 3) = 111$.

\begin{table*}[bthp]
\centering
\caption{
The nutrient input patterns in the data set of microbial community experiment, together with the basic characteristics of the soils under those conditions in the earliest (3days) and latest (24 days) time. The  population-balance-based similarity ${\cal B}$ and the OTU composition-based similarity ${\cal C}$ to the average of control communities are also shown.
The diversity and the dominance here represent the Shannon entropy of RSA: $S_\alpha = \sum_{r} - p_\alpha(r) \ln p_\alpha (r)$ and the proportion of the sum of the abundances of top 10 OTUs to the total abundance, respectively.
$\bar{\cal B}^{N_B}$ and $\bar{\cal C}^{N_C}$ are the averages of the proposed similarity to the control samples in the population balance and in the OTU composition which are defined as Eq. (\ref{eq_B}) and as Eq. (\ref{eq_C}), respectively.
}
\begin{tabular}{lccccrccccccc}
\midrule
\multicolumn{5}{c}{Condition} & \multicolumn{7}{c}{Characteristics}\\
\cmidrule(lr){1-5}
\cmidrule(lr){6-12}
$t$ (treatment) & Nitrogen & Carbon & Timing & $d$ (obs. day) &  Diversity $S$& Dominance & $\#_{\mbox{OTU}}$ & $\bar{\cal B}^{1000}$ & $\bar{\cal C}^{30}$ & $\bar{\cal C}^{100}$ & $\bar{\cal C}^{300}$ \\
\midrule
Control & none & none & - & 3 & 6.648 & 0.146 & 3805 & 0.969 & 0.804 & 0.790 & 0.749 \\ 
&&&& 24 & 6.823 & 0.112 & 4324 & 0.985 & 0.844 & 0.824 & 0.773 \\ 
\\
AG1 & \ce{NH4Cl} & Glucose & First Day & 3 & 2.588 & 0.875 & 1696 & 0.580 & 0.258 & 0.332 & 0.437 \\ 
&&&& 24 & 4.649 & 0.522 & 2861 & 0.708 & 0.175 & 0.064 & 0.257 \\
\\
AG4 &  \ce{NH4Cl} & Glucose & Weekly $\times 4$ &3 & 4.032 & 0.625 & 3162 & 0.568 & 0.351 & 0.436 & 0.542 \\  
&&&& 24 & 4.655 & 0.526 & 2906 & 0.737 & 0.104 & 0.155 & 0.310 \\ 
\\
AC1 &  \ce{NH4Cl} & Cellobiose & First Day & 3 & 2.678 & 0.864 & 1867 & 0.601 & 0.230 & 0.349 & 0.428 \\ 
&&&& 24 & 4.429 & 0.543 & 2462 & 0.726 & 0.070 & 0.008 & 0.220\\ 
\\
AC4 & \ce{NH4Cl} & Cellobiose & Weekly $\times 4$& 3 & 4.073 & 0.634 & 2897 & 0.592 & 0.353 & 0.494 & 0.588 \\ 
&&&& 24 & 5.012 & 0.433 & 3032 & 0.737 & -0.017 & 0.003 & 0.218 \\ 
\\
UG1 & Urea & Glucose & First Day& 3 & 3.302 & 0.774 & 2034 & 0.644 & 0.159 & 0.295 & 0.424 \\ 
&&&& 24 & 4.681 & 0.503 & 2668 & 0.736 & 0.085 & 0.058 & 0.227 \\ 
\\
UG4 & Urea & Glucose & Weekly $\times 4$ & 3 & 3.792 & 0.682 & 2652 & 0.584 & 0.347 & 0.517 & 0.575 \\ 
&&&& 24 & 5.155 & 0.407 & 2730 & 0.779 & 0.101 & 0.018 & 0.181 \\ 
\\
UC1 & Urea & Cellobiose & First Day & 3 & 3.615 & 0.737 & 2325 & 0.650 & 0.174 & 0.340 & 0.433 \\ 
&&&& 24 & 4.958 & 0.434 & 2621 & 0.846 & -0.016 & -0.161 & 0.121 \\ 
\\
UC4 & Urea & Cellobiose & Weekly $\times 4$ & 3 & 0.546 & 3.460 & 2582 & 0.560 & 0.295 & 0.501 & 0.547 \\ 
&&&& 24 & 0.690 & 4.828 & 2612 & 0.698 & 0.017 & 0.007 & 0.182 \\
 \bottomrule
\end{tabular}
\label{table_RSA}
\end{table*}

\begin{figure*}[thbp]
\includegraphics[width=0.45\linewidth]{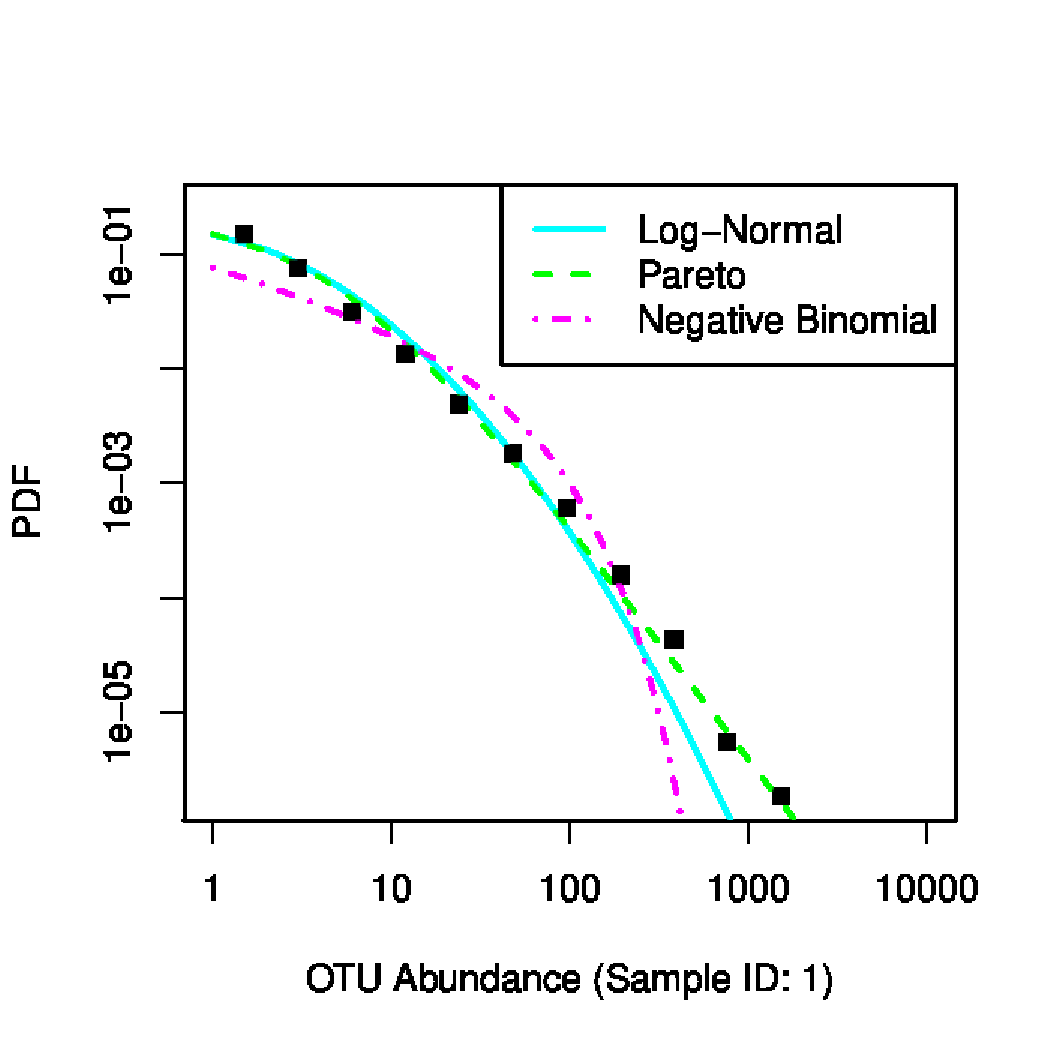}
\includegraphics[width=0.45\linewidth]{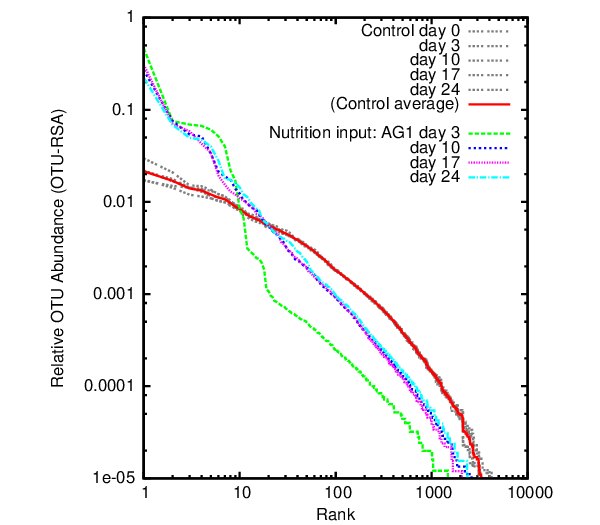}
\caption{
(Left)
An example of the probability distribution function of OTU abundance of soils before nutrition inputs (``control'', sample ID 1). The distribution is well fit by log-normal distribution and Pareto distribution.
(Right)
Rank-size plots of relative OTU abundances (OTU-RSA) of control soils and that of a soils after a one-shot nutrition input (AG1) in log-log scale.
While OTU-RSA shape is well kept among control soils, OTU-RSA of the soils after the perturbation are deformed significantly which is characterized by the enhanced dominance of top $\sim 10$ OTUs. OTU-RSAs in the later time stage relax to a smoother distribution, which is similar to the original shape.
}
\label{fig_RSAs}
\end{figure*}

\subsection*{Species Abundance Distributions}
We first confirm the basic characters of the control communities by its relative abundance distribution (RSA) of operational taxonomy units (OTU), which we shall call OTU-RSA.
For this purpose, we compare the fittings by classical theoretical distributions for species abundance of ecological systems, namely, log-normal distribution, power-law (Zipf/Pareto) distribution, and the negative binomial distribution.
An example of fitting is shown in the left panel of Fig. \ref{fig_RSAs}.
OTU-RSA of control soils are found to be better fit by log-normal distribution and power law distributions than by the negative binomial distribution, with a distinctive difference in the log-likelihood and equivalently in AIC, since all these distributions have two parameters.
The obtained fitting parameter values consistently indicate that the OTU-RSA of the soil microbe ecosystem before any disturbance is broad (see Materials and Methods and SI for detail).
This feature is consistent with the systems under natural environment\cite{Shoemaker2017}, implying that the control soils are keeping such an intact state.

The abundance distributions of the soils under or after nutrition input are also fat-tailed, with typically narrower shape comparing to that of control soils (Fig. \ref{fig_RSAs}, right panel).
A typical response of RSA to the nutrition input is characterized by an intial decrease of diversity
which is measured by the number of detectable OTUs or by Shannon entropy of OTU abundances
\begin{equation}
	S = - \sum_r p(r) \ln p(r),
\end{equation}
and an increase of the dominance of top tens of OTUs.
This decrease is followed by the recovery to a smoother distribution which is nearer to that of original control soils (TABLE \ref{table_RSA} and Supplemental Information).
Such changes in RSA clearly illustrates a homeostatic aspect of the soil microbial ecosystem: Fertilization typically first leads to more oligopolistic abundance distribution but the system later shows a resilience toward the original population balance.

\subsection*{Similarity in RSA shape}
The above observation about the typical change in the shape of OTU-RSAs naturally motivates us to define a similarity measure solely based on that.
A simple and natural choice is to take the linear correlation coefficient between the relative OTU abundances of soil $\alpha$ and soil $\beta$ in the same rank $r$,
\begin{equation}
    {\cal B}_{\alpha \beta}^{R_B}
	\equiv
	\frac{\displaystyle \sum_{r=1}^{R_B} \left( p^\alpha_r - \overline{p^\alpha} \right) \left( p^\beta_r - \overline{p^\beta} \right)}
	{\displaystyle \sqrt{ \sum_{r=1}^{R_B} \left( p^\alpha_r - \overline{p^\alpha} \right)^2} \sqrt{ \sum_{r=1}^{R_B} \left( p^\beta_r - \overline{p^\beta} \right)^2 }},
	\label{eq_B}
\end{equation}
where $p^\chi_r$ represents the relative abundance of the OTU whose abundance rank is $r$ in soil $\chi$, and $\overline{p^\chi} = \left. \sum_{r=1}^{R_B} p^\chi_r \right/ R_B$ denotes the average abundance up to the maximum rank $R_B$ to be taken into account. Since OTU-RSA has broad shape, the maximum rank $R_B$ in the following will be set at $1,000$ to include the information from as diverse OTUs as possible.

\begin{figure}[bthp]
\begin{center}
\includegraphics[width=0.7\linewidth]{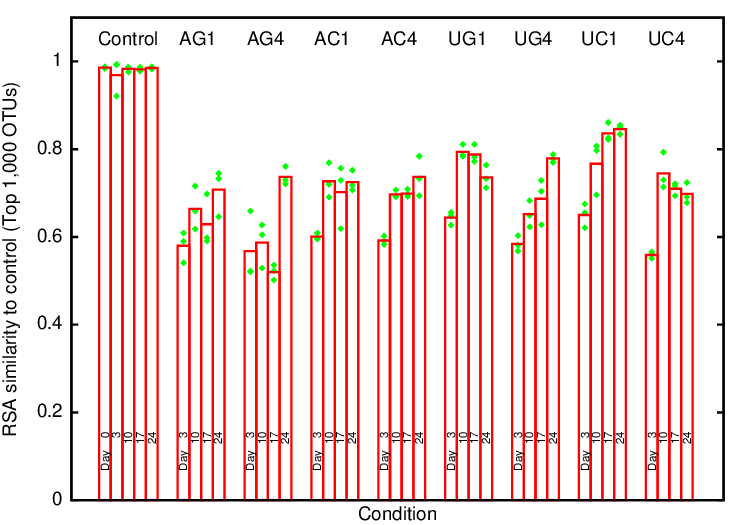}
\caption{
Similarities of the soil communities under the various conditions measured by the RSA-shape-based metric ${\mathcal B}$, against the communities of control soils.
Bars represent the average similarity of the samples under each treatment condition $t$ and the observation day $d$ to the control soils and the points represent the average similarity of each sample to the control soils.
While the control samples keep high similarity over time, ${\mathcal B}$ of the communities after the pulsive nutrition inputs show sharp drop and that is followed by partial recovery in the later time.
The similar dynamics of drop and recovery is observed also under successive nutrition inputs.}
\label{fig_RSAsim1000}
\end{center}
\end{figure}

The average ${\cal B}$ similarities of the soil communities under each treatment condition $t$ observed on day $d$ to the control soil,
\begin{equation}
	\bar{\cal B}(t, d) = \frac{\displaystyle \sum_{t_\alpha = t, d_\alpha = d; \ t_\beta = {\rm Control}} \cal{B}_{\alpha \beta} }{\displaystyle \sum_{t_\alpha = t, d_\alpha = d; \ t_\beta = {\rm Control}} 1},
\end{equation}
are shown in Fig. \ref{fig_RSAsim1000} (bars).
We can see that the temporal evolutions of the soils in this measure against various inputs (different combinations of nutrition and the different schedules) universally show a homeostatic response, i.e. the similarity to the original control community first drops and that recovers later on.
The response of the soils against the continuous inputs is particularly interesting, because the recovery takes place during the successive nutrition input. This is why we regard it homeostatic, rather than a mere relaxation back to the original state under absence of the perturbation.

\subsection*{Similarity in OTU abundance rank orders}
We next seek a measure to capture a totally different aspect of the soil microbial community.
Because the systems have been confirmed to show robustness in its population {\it balance} (the shape of RSA), the remaining degree of freedom in OTU abundance distribution is in its {\it composition}.
Suppose that two communities have the same OTU-RSA, then yet two communities can be different if OTUs occupying each rank position are different each other.
In other words, the order of OTUs in relative abundance (whether a particular OTU is major to another particular OTU) is independent of the shape of OTU-RSA distribution.

Our similarity measure for the composition of OTUs, with respect to the OTU-RSA information, is defined by the Kendall's rank correlation between the abundance ranks of top $N_C$ major OTUs of the two communities in comparison, $\alpha$ and $\beta$:
\begin{equation}
    {\mathcal C}_{\alpha \beta}^{N_C}
	\equiv	
	\frac{\#_c - \#_d}{N_C (N_C -1)/2},
	\label{eq_C}
\end{equation}
where $\#_c$ and $\#_d$ represent the numbers of concordant OTU rank pairs, i.e.
$\large( r^\alpha_i - r^\alpha_j \large) \large( r^\beta_i - r^\beta_j \large) > 0,$
and discordant OTU rank pairs, i.e.
$\large( r^\alpha_i - r^\alpha_j \large) \large( r^\beta_i - r^\beta_j \large) < 0,$
respectively.
Where $\displaystyle r^\chi_k$ represents the abundance rank of OTU $k$ in the soil $\chi$. If a certain OTU is not detected in the sample, we replace $\displaystyle r^\gamma_k$ by a uniform number $r_{\max}$ which is larger than the total number of OTUs. Therefore, for example, if both OTUs $i$ and $j$ are missing in a soil $\alpha$, any combination of OTU abundance ranks in the other soil $\beta$ ($r_i^\beta$ and $r_j^\beta$) will not be counted as concordant or discordant.
The parameter $N_C$ is chosen to be large enough to take as much information as possible and not too large to avoid an error from the detection limit.
To determine proper $N_C$, we first check how many OTUs are commonly observed in the soils under different conditions.
For this purpose, we evaluate the overall popularity (majority) of each OTU by aggregating the all OTU abundances across the available conditions.  
We find that almost all of top $300$ major OTUs in the aggregated data are found to be shared in all soils (shown in SI, Fig. \ref{fig_Simpson}).
Some of the OTUs in the following overall majority rank range becomes to be absent (undetectable) in some soils.
Therefore, in the following we choose $N_C = 100$, OTUs up to this are expected to provide the majority-minority information well above the detection limit of OTU abundances in the current experiment.
One can confirm that the following results are not sensitive to the precise choice of $N_C$ (for example, for $N_C = 30$ and $N_C = 300$).

The present measure ${\mathcal C}$ takes $1$ if the rank orders between OTUs (major-minority relations) is kept for all pairs, and it takes a value near $0$ if there is no correlation.
It can take negative value if the reversal in the majority order is dominant, ultimately down to $-1$ when all the majority order pair are flipped.
Because this measure is independent of the shape of RSA distribution, and also because the swap in the ordering of OTU abundances does not change the RSA shape, the previously defined ``population-balance measure'' ${\cal B}$ and the currently introduced ``composition measure'' ${\cal C}$ can be said to be independent, or {\it orthogonal}, to each other.

\begin{figure}[bthp]
\begin{center}
\includegraphics[width=0.7\linewidth]{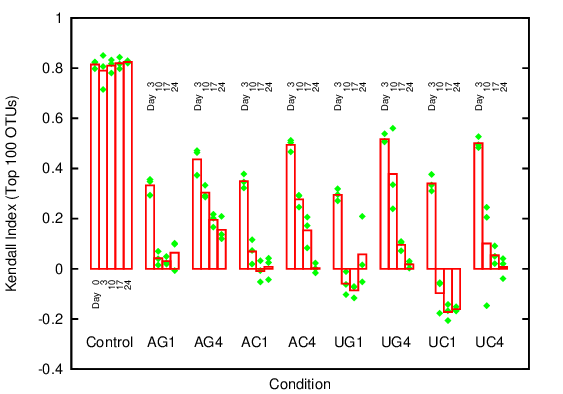}
\caption{
OTU composition similarities ${\cal C}$ to the control soils. Each bar represents the sample average of the soils under each perturbation condition and timing, with the similarity of each soil is shown with the symbols (three samples for each condition).
Compared to the high similarity among control conditions, ${\cal C}$ similarities after first fertilization (Day 3) are lower. And the similarities show further decrease in the later time, even for the pulsive input, i.e. during the period without any further input.
}
\label{fig_Kendall100}
\end{center}
\end{figure}

The OTU composition similarities ${\mathcal C}$ measured using the top $N_C = 100$ major OTUs are shown in Fig.\ref{fig_Kendall100}, where bars again represent the average similarity of the samples with treatment condition $t$ and the observation day $d$ to the control soils:
\begin{equation}
	\bar{\cal C}(t, d) = \frac{\displaystyle \sum_{t_\alpha = t, d_\alpha = d; \ t_\beta = {\rm Control}} \cal{C}_{\alpha \beta} }{\displaystyle \sum_{t_\alpha = t, d_\alpha = d; \ t_\beta = {\rm Control}} 1}.
\end{equation}
The similarity among the control soil is confirmed to be high in this metric. Compared to that, the average similarities of the soils 3 days after nutrition inputs to the control soil is low, meaning that the composition of OTUs are changed greatly from the control soils.
What is remarkable is that the similarity of the perturbed soils to the control soils shows further decrease in the later time.
This tendency is universally observed for different conditions, and for the the choice of $N_C$ as shown in the Table I and SI.
Recalling that the ${\mathcal B}$ similarity shows an increase in the same moment of later time stage, the observed further change in OTU composition takes place during the recovery process of the population balance.
This implies that the response in the composition measure ${\mathcal C}$ is not just due to a slower response, but is capturing the plasticity of the soil community to the nutrition input.

\section*{Discussion}
In this study, we have proposed simple similarity measures to characterize the response of the soil microbial communities to nutrition inputs.
The first one, ${\mathcal B}$, is based on the {\it balance} of the populations (shape of OTU-RSA), and the other one, ${\mathcal C}$, is based on the {\it composition} of OTUs.
The both similarities among the control soil communities are kept during our observation time of 24 days, meaning that the stationary character of the communities under no external inputs is well captured by these macroscopic measures.

When the nutrition inputs are applied, the immediate response of the communities appears as the significant drops in both similarities.
The typical temporal evolutions of the similarities in the later time are, however, different each other.
The  ${\mathcal B}$ similarities of the perturbed soils to the original communities universally show a recovery in time. The recovery is observed even under the continuous nutrition input.
Therefore this measure universally characterizes the robust (or homeostatic) nature of the soil communities.
In ${\mathcal C}$, on the other hand, an opposite character of the soil's response is detected. The ${\mathcal C}$ of the perturbed soils generally show further decrease in time, in the middle of the recovery period in ${\mathcal B}$ similarity (10, 17, and 24 days after the onset of nutrition inputs).
This further decrease is universal among different nutrients and input schedule, while the continuous input tends to results in more monotonic decay.
Therefore the plastic (or adaptive) aspect of the soil communities is also well captured by the ${\mathcal C}$ similarity.

Most of the attempts which have so far been made to disentangle the dynamics of complex microbial community structures rely on beta-diversity metrics between spatially/temporally distant communities. The concept of beta-diversity appears to be two-fold. While it was originally proposed as a between-community difference in the presence/absence of each species \cite{whittaker1960vegetation}, currently popular metrics such as Bray-Curtis \cite{bray1957ordination} and weighted UniFrac distances \cite{lozupone2007quantitative} bear the information on the quantitative balances between community members.
The two indices we have proposed here are approximately congruent with these beta-diversity-based distances, in the sense that both ${\mathcal B}$ and ${\mathcal C}$ are reflected.
This can be seen in the present data that the drastic changes the soil bacterial communities underwent after the initial perturbations (i.e. between days $0$ and $3$) is visible by the beta-diversity measures (Fig. \ref{fig_NMDS} in SI). However, in these measures, the rapid recovery in species balances (i.e. change in ${\mathcal B}$) has overwhelmed and masked the long-lasting effect on the change in microbial community members (i.e. change in ${\mathcal C}$). It is also notable that the contrast between two schemes of nutritional input schedule, one-time and four times, are obscure in ${\mathcal B}$ (and in canonical beta-diversity) but clearly highlighted by ${\mathcal C}$.
Therefore, by decomposing the canonical beta-diversities into two features by ${\mathcal B}$ and ${\mathcal C}$, otherwise mingled differences between incubation conditions are better illuminated.

For the understandings of the robustness and plasticity of interacting communities in more general sense, present discovery can be regarded as a good guiding principle.
Let us consider trying the construction of a dynamical community model, under a constraint that which can explain the coexistance of observed robustness and plasticity. As is evident from the large change in OTU-RSA shape, the homeostatic response to the nutrition input we have observed by ${\mathcal B}$ is beyond linear (local) stability argument. Therefore, the observed trajectory should be explained as a transition from an attractor (a stable state) to another attractor triggered by the change of nutrition parameter or a large disturbance in population distribution. A challenge in such approach is that one should reproduce the order of the response: the recovery in OTU balance takes place first and the dynamics of the OTU composition change is slower.
Another possibility to explain the robustness in OTU-RSA, with keeping the degree of freedom of change in the OTU composition, would be a more statistical (entropic) mechanism under some hidden microbial constraints. The neutral theory is a good candidate for this, though it is not straightforward to test that for the observed response. This is because the environment in this study is dynamically changing and hence the key parameters in the framework, such as the birth and death rate, dispersion, and so on of each OTU are not kept. And even if the effect of nutrition input is nicely modeled by the abrupt change in such parameters, to reproduce the responses in the later time remains as a difficult task.

Finally, present finding is also meaningful in the agricultural context. It is important in agriculture to put organic matter such as compost into the soil to maintain the quality of the soil. Fertilizers as well are basically essential for producing crops on agricultural field soil. And, giving carbon source and/or nitrogen source to soil in this way can be regarded as one of the disturbance of soil ecosystem. It was reported that fertilization had a greater effect on soil bacterial community structure than crop rotation and growth stage (Guo et al. 2020). It is also shown that soil organic carbon and total nitrogen are two of the major determinants of community composition (Li et al. 2017). The disturbance of the shape of the OTU-RSA curve for the soil amended with carbon and nitrogen sources (glucose and ammonium chloride, respectively) (Fig. 1) clearly indicate a rapid changes in bacterial community structure. In addition, when combined with traditional analyses such as non-metric multidimensional scaling (NMDS) (SI Fig. 4), the overall picture of the changes that occur in the community structure becomes easier to understand.

\section*{Methods}
\subsection*{Sample Preparation and Primary Sequence Analysis}
Nucleotide sequence files registered in DDBJ/ENA/GenBank under the accession numbers DRR157393-157518 (DRA007564) and DRR169428-169475 (DRA008081) were retrieved. These sequences were generated by amplicon sequencing targeting 16S rRNA genes in soil microcosms amended with labile carbon and nitrogen sources (partly described in \cite{Mise2020}). These genes are the commonly-used "marker" for the identification of prokaryotes. The soil microcosm incubation experiment was designed in a full-factorial manner: besides control samples without nutritional amendment, eight treatment groups, differing in carbon source (glucose or cellobiose), nitrogen source (ammonium chloride or urea), and timing(s) of nutritional amendment (weekly amendment over the experiment period or abrupt amendment of quadruple amount at the beginning of incubation), were prepared (Table 1). The total amount of carbon and nitrogen input was the same in all groups. Twelve microcosms were constructed for each group, three of which were destructively sampled after 3, 10, 17, and 24 days of incubation. Additionally, the pre-incubation (i.e. day 0) soil was examined in triplicate. Sampled soil was subjected to total DNA extraction, purification, amplification of partial 16S rRNA gene sequence, and Illumina high-throughput sequencing as described previously \cite{Mise2020}. 

Low-quality (i.e. expected errors of 0.5 bases or more) sequences were removed using USEARCH v9.2.64 \cite{EdgarUSEARCH}. Subsequently singleton sequences were removed, and the remaining sequences were clustered into operational taxonomic units (OTUs) at a similarity threshold of 97\% or more using UPARSE \cite{EdgarUPARSE}. The OTU representative sequences (i.e. most frequently-observed sequence within those clustered into the OTU) were subjected to UCHIME2 \cite{EdgarUCHIME2} to filter out chimera-like OTUs. Finally, all the quality-filtered sequences, including singletons, were mapped to OTU representative sequences using USEARCH v9.2.64. To detect organelle-like OTUs, the OTU representative sequences were taxonomically annotated by RDP classifier \cite{RDP2007} trained with Greengenes 13\_8 clustered at 97\% \cite{Greengenes}. OTUs annotated as "family mitochondria" or "class Chloroplast" were precluded from further analyses.

\subsection*{Species Abundance Distributions}
To characterize the SAD, we compare the fittings by log-normal distribution
\begin{equation}
P^{LN} (\mu, \sigma; x)
=
\frac{1}{\displaystyle  \sigma x \sqrt{2\pi}} \exp\left\{-\frac{\displaystyle (\ln x - \mu)^2}{\displaystyle 2\sigma^2} \right\},
\label{eq_pLN}
\end{equation}
power-law (Zipf/Pareto) distribution
\begin{equation}
P^{PL} (\alpha, \theta; x) = \frac{\alpha \theta^\alpha}{(x+\theta)^{\alpha+1}},
\label{eq_pPL}
\end{equation}
and the negative binomial distribution
\begin{equation}
P^{NB} (s, p; x) = \frac{\Gamma(x+s)}{\Gamma(x+1) \ \Gamma(s)} \ p^s (1-p)^x,
\label{eq_pNB}
\end{equation}
where $x$ denote the OTU abundance.

The obtained fitting parameter values;
$ \ln \mu \sim \ln \sigma \sim 1.5$ for the log-normal distribution,
shape parameter $1.0 < \alpha < 1.5$ for Pareto distribution,
and the size $0 < s < 0.75$ and the mean $p \sim 0.01$ for the negative binomial distribution,
consistently indicate that the OTU-SAD of the soil microbe ecosystem before any artificial nutrition input is very broad, i.e. critical or fat-tailed (see SI for detail).

\hspace{1cm}
\section*{Acknowledgement}
TS was partly supported by JSPS KAKENHI grant number JP18K03449 and JP23K03256.
TS and SO were partly supported by JSPS KAKENHI grant number JP17K19220.
KMo was partly supported by JSPS KAKENHI Grant Number JP20K19885.

\bibliography{ecosoil.bib}

\clearpage
\renewcommand{\thefigure}{S-\arabic{figure}}
\setcounter{figure}{0}
\renewcommand{\thetable}{S-\Roman{table}}
\setcounter{table}{0}

\section*{Supplementary Information}

\subsection*{NMDS}
\begin{figure}[bthp]
\includegraphics[width=0.5\linewidth]{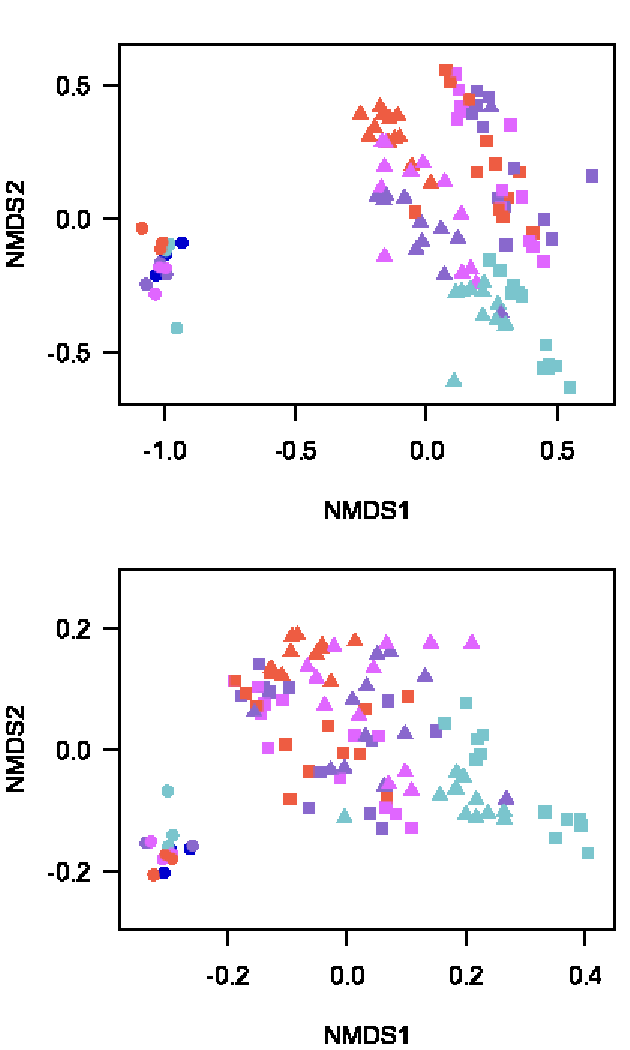}
\caption{A non-metric dimensional scaling plot denoting the changes in microbial community compositions based on Bray-Curtis dissimilarities (upper) and weighted UniFrac distances (bottom).
Circle, square, and triangle points denote no, abrupt, and weekly amendments of nutrients, respectively.
Colors of the points corresponds to the stages (time after treatment):  light blue, dark purple, light purple, and red represent the earliest, second earliest, third earliest, and the latest stages, respectively.
}
\label{fig_NMDS}
\end{figure}

\clearpage
\subsection*{${\mathcal B}$ similarity to control soils}
\begin{table*}[h]
\centering
\caption{
The average population-balance-based similarity ${\cal \bar{B}}$ to the control soils, with other characteristics relating to the diversity and RSA shape.
${\cal B}$ similarity is defined as Eq. (\ref{eq_B}), with $R = 1,000$.
The diversity and the dominance here represent the Shannon entropy of RSA: $S^\alpha = \sum_{r^\alpha} - p_r^\alpha \ln p_r^\alpha$ and the proportion of the sum of the abundances of top $10$ OTUs to the total abundance, respectively.
}
\begin{tabular}{lr|rrr|rr|rrr}
Condition & Day &
Diversity & Dominance & $\#_{\mbox{OTU}}$ &
$\bar{\cal B}^{100}$ & $\bar{\cal B}^{1000}$ & $\bar{\cal C}^{30}$ & $\bar{\cal C}^{100}$ & $\bar{\cal C}^{300}$\\
\midrule
Control & 0 & 6.741 & 0.128 & 4293 & 0.978 & 0.986 & 0.843 & 0.814 & 0.776\\ 
 & 3 & 6.648 & 0.146 & 3805 & 0.966 & 0.969 & 0.804 & 0.790 & 0.749 \\ 
 & 10 & 6.689 & 0.124 & 3607 & 0.973 & 0.983 & 0.832 & 0.810 & 0.761 \\ 
 & 17 & 6.692 & 0.124 & 3659 & 0.970 & 0.982 & 0.834 & 0.819 & 0.766 \\ 
 & 24 & 6.823 & 0.112 & 4324 & 0.980 & 0.985 & 0.844 & 0.824 & 0.773 \\ 
\midrule
AG1 & 3 & 2.588 & 0.875 & 1696 & 0.700 & 0.580 & 0.258 & 0.332 & 0.437 \\ 
 & 10 & 4.373 & 0.567 & 3080 & 0.756 & 0.664 & 0.224 & 0.042 & 0.230 \\ 
 & 17 & 4.185 & 0.583 & 2761 & 0.721 & 0.623 & 0.157 & 0.031 & 0.223 \\ 
 & 24 & 4.649 & 0.522 & 2861 & 0.793 & 0.708 & 0.175 & 0.064 & 0.257 \\
\midrule
AG4 & 3 & 4.032 & 0.625 & 3162 & 0.665 & 0.568 & 0.351 & 0.436 & 0.542 \\ 
 & 10 & 4.814 & 0.476 & 3399 & 0.666 & 0.587 & 0.304 & 0.304 & 0.404 \\ 
 & 17 & 3.985 & 0.595 & 2954 & 0.613 & 0.520 & 0.207 & 0.195 & 0.304 \\ 
 & 24 & 4.655 & 0.526 & 2906 & 0.828 & 0.737 & 0.104 & 0.155 & 0.310 \\ 
\midrule
AC1 & 3 & 2.678 & 0.864 & 1867 & 0.728 & 0.601 & 0.230 & 0.349 & 0.428 \\ 
 & 10 & 4.062 & 0.644 & 2600 & 0.833 & 0.727 & 0.089 & 0.069 & 0.272 \\ 
 & 17 & 4.249 & 0.585 & 2540 & 0.796 & 0.702 & 0.043 & -0.009 & 0.221 \\ 
 & 24 & 4.429 & 0.543 & 2462 & 0.812 & 0.726 & 0.070 & 0.008 & 0.220 \\ 
\midrule
AC4 & 3 & 4.073 & 0.634 & 2897 & 0.697 & 0.592 & 0.353 & 0.494 & 0.588 \\ 
 & 10 & 4.391 & 0.588 & 2856 & 0.804 & 0.697 & 0.119 & 0.277 & 0.460 \\ 
 & 17 & 3.977 & 0.660 & 2665 & 0.814 & 0.699 & 0.048 & 0.153 & 0.325 \\ 
 & 24 & 5.012 & 0.433 & 3032 & 0.804 & 0.737 & -0.017 & 0.003 & 0.218 \\ 
\midrule
UG1 & 3 & 3.302 & 0.774 & 2034 & 0.764 & 0.644 & 0.159 & 0.295 & 0.424 \\ 
 & 10 & 4.622 & 0.510 & 2475 & 0.877 & 0.794 & 0.080 & -0.059 & 0.167 \\ 
 & 17 & 4.786 & 0.469 & 2522 & 0.860 & 0.788 & 0.070 & -0.086 & 0.143 \\ 
 & 24 & 4.681 & 0.503 & 2668 & 0.820 & 0.736 & 0.085 & 0.058 & 0.227 \\ 
\midrule
UG4 & 3 & 3.792 & 0.682 & 2652 & 0.693 & 0.584 & 0.347 & 0.517 & 0.575 \\ 
 & 10 & 4.073 & 0.638 & 2601 & 0.762 & 0.652 & 0.232 & 0.378 & 0.454 \\ 
 & 17 & 4.417 & 0.556 & 2504 & 0.781 & 0.687 & 0.089 & 0.095 & 0.234 \\ 
 & 24 & 5.155 & 0.407 & 2730 & 0.840 & 0.779 & 0.101 & 0.018 & 0.181 \\ 
\midrule
UC1 & 3 & 3.615 & 0.737 & 2325 & 0.765 & 0.650 & 0.174 & 0.340 & 0.433 \\ 
 & 10 & 4.542 & 0.528 & 2501 & 0.853 & 0.767 & 0.007 & -0.097 & 0.147 \\ 
 & 17 & 4.909 & 0.447 & 2534 & 0.904 & 0.836 & -0.012 & -0.172 & 0.109 \\ 
 & 24 & 4.958 & 0.434 & 2620 & 0.913 & 0.846 & -0.016 & -0.161 & 0.121 \\ 
\midrule
UC4 & 3 & 3.460 & 0.729 & 2582 & 0.670 & 0.559 & 0.295 & 0.501 & 0.547 \\ 
 & 10 & 4.690 & 0.517 & 2863 & 0.835 & 0.745 & 0.038 & 0.101 & 0.285 \\ 
 & 17 & 4.634 & 0.519 & 2518 & 0.798 & 0.710 & -0.009 & 0.054 & 0.226 \\ 
 & 24 & 4.828 & 0.465 & 2612 & 0.770 & 0.698 & 0.017 & 0.007 & 0.182 \\
 \bottomrule
\end{tabular}
\label{table_SimilaritiesInDetail}
\end{table*}

\clearpage
\subsection*{Similarity of the control soils in OTU composition}
\begin{figure}[bthp]
\includegraphics[width=0.5\linewidth]{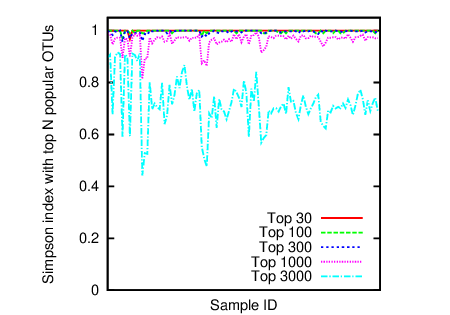}
\caption{The overlap coefficient (Szymkiewicz-Simpson coefficient)
$ S_{\alpha \beta} \equiv \frac{|\alpha \cap \beta|}{\min(|\alpha|, |\beta|)}$ between the set of top $N$ popular OTUs $\alpha$ and the OTUs in each soil $\beta$. Almost all of top $300$ major OTUs are detected in all soils while some of top $1{,}000$ popular OTUs are missing (i.e. absent or undetectable).
}
\label{fig_Simpson}
\end{figure}

\clearpage
\subsection*{Characteristics of RSA distributions}
\begin{figure*}[hbtp]
\includegraphics[width=0.5\linewidth]{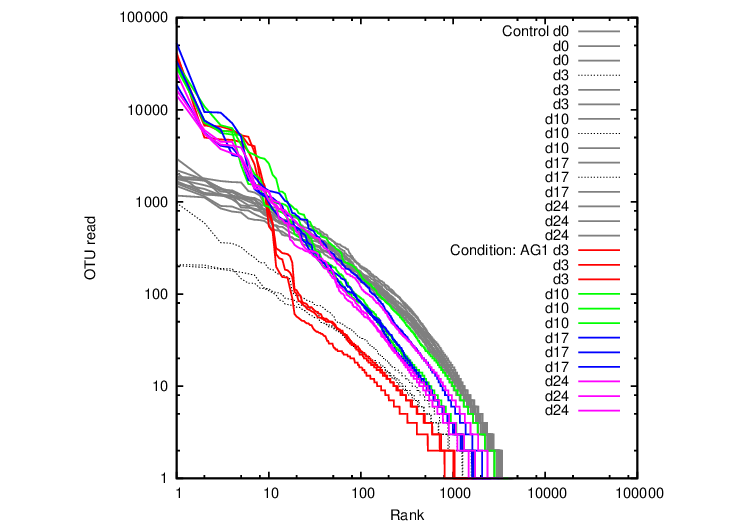}
\includegraphics[width=0.5\linewidth]{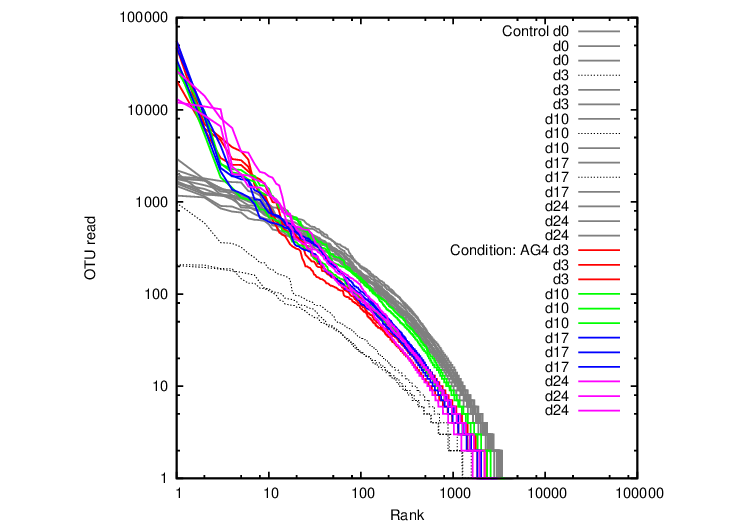}
\caption{Rank-size distributions of OTUs under fertilization with Ammonium chloride and Glucose (Left: abrupt, Right: weekly 4 times).}
\label{sppl_fig_SAD_BC}
\end{figure*}

\begin{figure*}[hbtp]
\includegraphics[width=0.5\linewidth]{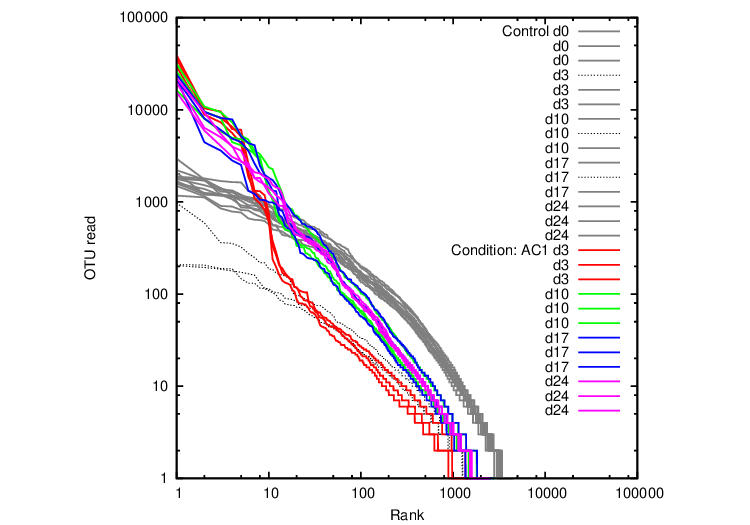}
\includegraphics[width=0.5\linewidth]{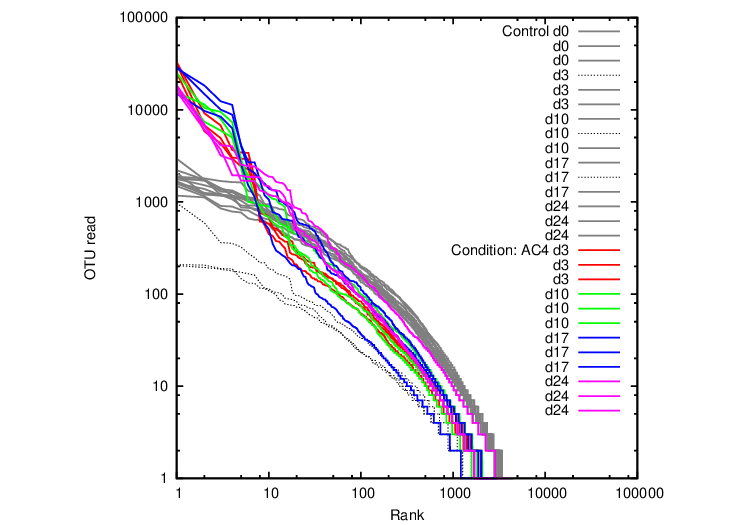}
\caption{Rank-size distributions of OTUs under fertilization with Ammonium chloride and Cellobiose (Left: abrupt, Right: weekly 4 times).}
\label{sppl_fig_SAD_DE}
\end{figure*}

\begin{figure*}[hbtp]
\includegraphics[width=0.5\linewidth]{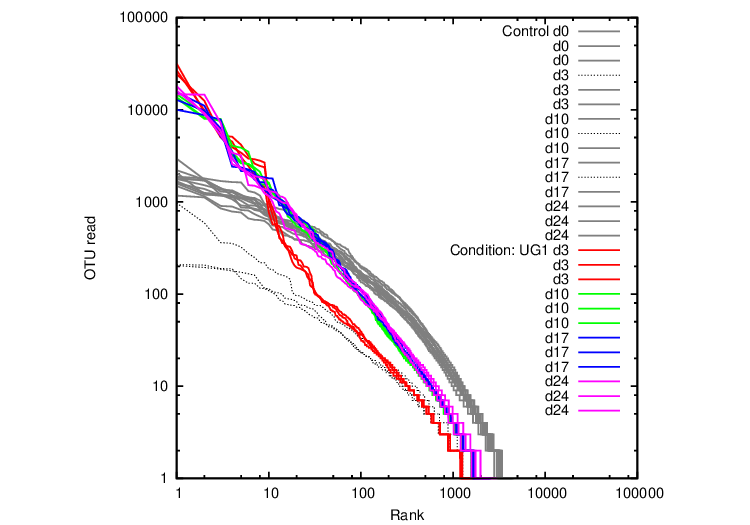}
\includegraphics[width=0.5\linewidth]{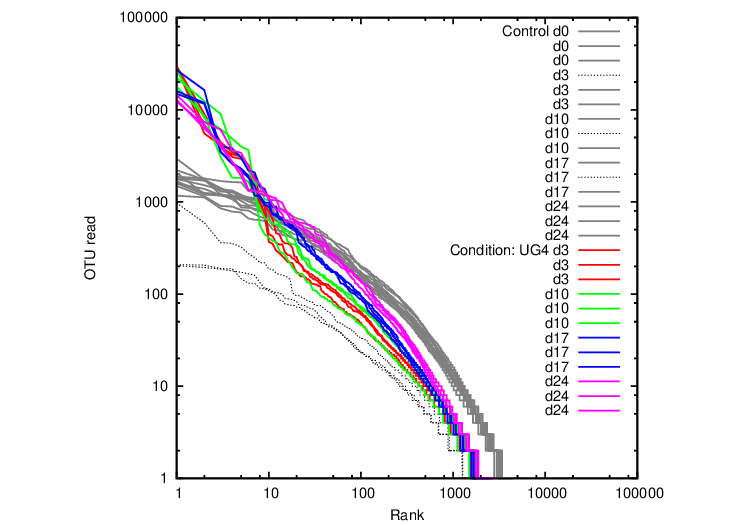}
\caption{Rank-size distributions of OTUs under fertilization with Urea and Glucose (Left: abrupt, Right: weekly 4 times).}
\label{sppl_fig_SAD_FG}
\end{figure*}

\begin{figure*}[hbtp]
\includegraphics[width=0.5\linewidth]{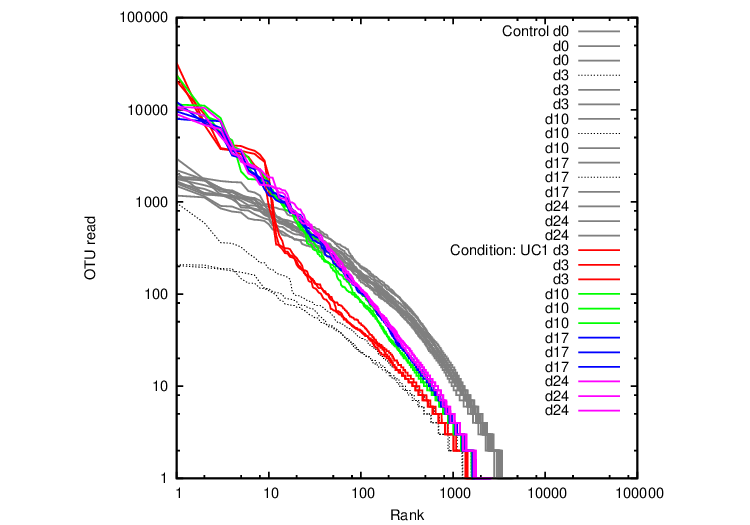}
\includegraphics[width=0.5\linewidth]{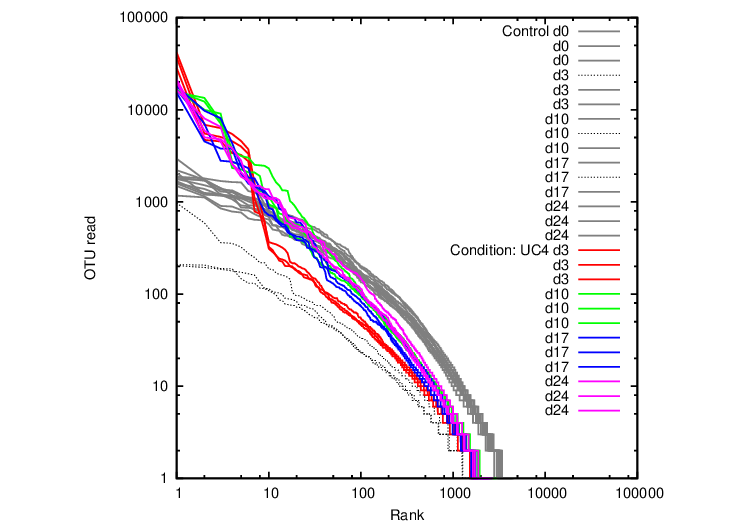}
\caption{Rank-size distributions of OTUs under fertilization with Urea and Cellobiose (Left: abrupt, Right: weekly 4 times).}
\label{sppl_fig_SAD_HI}
\end{figure*}

\begin{table}[bthp]
\centering
\caption{Characteristics of the rank-size plots of control soils.}
\begin{tabular}{rr|rrrr}
SampleID & Day & Diversity & Dominance & $\#_{\mbox{OTU}}$ & tot. $\#_{\mbox{read}}$\\
\midrule
1 & 0 & 6.740 & 0.126 & 4246 & 97138 \\ 
2 & 0 & 6.723 & 0.133 & 4213 & 79576 \\ 
3 & 0 & 6.759 & 0.125 & 4421 & 87402 \\ 
4 & 3 & 6.495 & 0.187 & 2696 & 20039 \\ 
6 & 3 & 6.736 & 0.120 & 4441 & 105427 \\ 
7 & 10 & 6.755 & 0.113 & 4304 & 90000 \\ 
8 & 10 & 6.647 & 0.119 & 2123 & 12961 \\ 
9 & 10 & 6.665 & 0.140 & 4395 & 92151 \\ 
10 & 17 & 6.713 & 0.125 & 4514 & 113768 \\ 
11 & 17 & 6.601 & 0.122 & 2124 & 13262 \\ 
12 & 17 & 6.760 & 0.124 & 4341 & 87653 \\ 
13 & 24 & 7.026 & 0.094 & 4602 & 83224 \\ 
14 & 24 & 6.727 & 0.119 & 4023 & 73965 \\ 
15 & 24 & 6.717 & 0.123 & 4347 & 88142 \\
\bottomrule
\end{tabular}
\end{table}

\begin{table}[bthp]
\centering
\caption{Characteristics of the rank-size plots of the soils with Ammonium chloride as the nitrogen source.}
\begin{tabular}{rr|rrrr}
SampleID & Day & Diversity & Dominance & $\#_{\mbox{OTU}}$ & tot. $\#_{\mbox{read}}$\\
\midrule
(AG1)\quad
16 & 3 & 2.277 & 0.902 & 1479 & 77263 \\ 
17 & 3 & 2.790 & 0.857 & 1819 & 84824 \\ 
18 & 3 & 2.698 & 0.866 & 1791 & 86315 \\ 
19 & 10 & 5.071 & 0.464 & 3956 & 146007 \\ 
20 & 10 & 4.004 & 0.619 & 2615 & 94330 \\ 
21 & 10 & 4.043 & 0.619 & 2669 & 111668 \\ 
22 & 17 & 4.486 & 0.530 & 2548 & 79529 \\ 
23 & 17 & 3.988 & 0.619 & 2567 & 97617 \\ 
24 & 17 & 4.079 & 0.600 & 3170 & 152920 \\ 
25 & 24 & 4.075 & 0.605 & 2333 & 84754 \\ 
26 & 24 & 4.636 & 0.537 & 2740 & 69740 \\ 
27 & 24 & 5.237 & 0.422 & 3511 & 98191 \\
\midrule
(AG4)\quad
28 & 3 & 4.807 & 0.515 & 3222 & 84241 \\ 
29 & 3 & 3.535 & 0.695 & 2933 & 103057 \\ 
30 & 3 & 3.753 & 0.664 & 3333 & 124647 \\ 
31 & 10 & 4.410 & 0.532 & 3053 & 86813 \\ 
32 & 10 & 5.019 & 0.451 & 3497 & 112981 \\ 
33 & 10 & 5.013 & 0.447 & 3647 & 122846 \\ 
34 & 17 & 4.020 & 0.599 & 3035 & 117447 \\ 
35 & 17 & 4.126 & 0.571 & 2816 & 87447 \\ 
36 & 17 & 3.809 & 0.615 & 3012 & 123010 \\ 
37 & 24 & 4.514 & 0.544 & 2541 & 81316 \\ 
38 & 24 & 4.404 & 0.571 & 2929 & 123503 \\ 
39 & 24 & 5.049 & 0.464 & 3250 & 87409 \\
\midrule
(AC1)\quad
40 & 3 & 2.862 & 0.837 & 2288 & 84250 \\ 
41 & 3 & 2.557 & 0.879 & 1711 & 89076 \\ 
42 & 3 & 2.616 & 0.875 & 1604 & 70229 \\ 
43 & 10 & 4.196 & 0.633 & 2519 & 84433 \\ 
44 & 10 & 3.823 & 0.689 & 2418 & 95165 \\ 
45 & 10 & 4.165 & 0.610 & 2863 & 132677 \\ 
46 & 17 & 4.093 & 0.591 & 2250 & 69308 \\ 
47 & 17 & 4.379 & 0.572 & 2796 & 121407 \\ 
48 & 17 & 4.275 & 0.592 & 2574 & 95467 \\ 
49 & 24 & 4.666 & 0.494 & 2517 & 81237 \\ 
50 & 24 & 4.212 & 0.579 & 2348 & 94279 \\ 
51 & 24 & 4.407 & 0.555 & 2523 & 84132 \\
\midrule
(AC4)\quad
52 & 3 & 4.077 & 0.633 & 2691 & 72891 \\ 
53 & 3 & 4.199 & 0.610 & 3037 & 91357 \\ 
54 & 3 & 3.942 & 0.659 & 2963 & 95568 \\ 
55 & 10 & 4.531 & 0.562 & 2810 & 75738 \\ 
56 & 10 & 4.153 & 0.636 & 2615 & 75640 \\ 
57 & 10 & 4.489 & 0.567 & 3145 & 110202 \\ 
58 & 17 & 3.997 & 0.660 & 2850 & 124981 \\ 
59 & 17 & 3.602 & 0.733 & 2117 & 63296 \\ 
60 & 17 & 4.333 & 0.586 & 3028 & 132373 \\ 
61 & 24 & 4.741 & 0.477 & 2635 & 84676 \\ 
62 & 24 & 5.466 & 0.364 & 3847 & 125750 \\ 
63 & 24 & 4.829 & 0.458 & 2614 & 83179 \\
\bottomrule
\end{tabular}
\end{table}

\begin{table}[hbtp]
\centering
\caption{Characteristics of the rank-size plots of the soils with Urea as the nitrogen source }
\begin{tabular}{rr|rrrr}
SampleID & Day & Diversity & Dominance & $\#_{\mbox{OTU}}$ & tot. $\#_{\mbox{read}}$\\
\midrule
(UG1)\quad
64 & 3 & 3.288 & 0.783 & 1968 & 78348 \\ 
65 & 3 & 3.384 & 0.755 & 2018 & 76670 \\ 
66 & 3 & 3.236 & 0.784 & 2116 & 85661 \\ 
67 & 10 & 4.608 & 0.513 & 2456 & 87498 \\ 
68 & 10 & 4.621 & 0.512 & 2480 & 93963 \\ 
69 & 10 & 4.637 & 0.505 & 2491 & 91091 \\ 
70 & 17 & 4.837 & 0.465 & 2530 & 90556 \\ 
71 & 17 & 4.869 & 0.453 & 2463 & 89096 \\ 
72 & 17 & 4.651 & 0.490 & 2573 & 95837 \\ 
73 & 24 & 4.789 & 0.470 & 2629 & 96630 \\ 
74 & 24 & 4.575 & 0.505 & 2406 & 91239 \\ 
75 & 24 & 4.680 & 0.535 & 2971 & 91612 \\
\midrule
(UG4)\quad
76 & 3 & 3.866 & 0.669 & 2714 & 78639 \\ 
77 & 3 & 3.861 & 0.674 & 2736 & 84624 \\ 
78 & 3 & 3.647 & 0.702 & 2506 & 72037 \\ 
79 & 10 & 3.558 & 0.733 & 2398 & 77141 \\ 
80 & 10 & 4.519 & 0.545 & 2632 & 69803 \\ 
81 & 10 & 4.140 & 0.637 & 2775 & 90771 \\ 
82 & 17 & 4.604 & 0.535 & 2540 & 87544 \\ 
83 & 17 & 4.531 & 0.536 & 2410 & 76750 \\ 
84 & 17 & 4.116 & 0.599 & 2564 & 96683 \\ 
85 & 24 & 5.138 & 0.411 & 2603 & 82586 \\ 
86 & 24 & 5.218 & 0.386 & 2762 & 89540 \\ 
87 & 24 & 5.110 & 0.424 & 2826 & 101649 \\
\midrule
(UC1)\quad
88 & 3 & 3.848 & 0.700 & 2330 & 72345 \\ 
89 & 3 & 3.626 & 0.741 & 2357 & 75994 \\ 
90 & 3 & 3.370 & 0.771 & 2288 & 87739 \\ 
91 & 10 & 4.679 & 0.504 & 2516 & 83231 \\ 
92 & 10 & 4.487 & 0.552 & 2437 & 89271 \\ 
93 & 10 & 4.460 & 0.527 & 2550 & 102115 \\ 
94 & 17 & 4.890 & 0.447 & 2509 & 90361 \\ 
95 & 17 & 4.902 & 0.456 & 2549 & 87501 \\ 
96 & 17 & 4.935 & 0.439 & 2545 & 87503 \\ 
97 & 24 & 4.987 & 0.431 & 2634 & 96957 \\ 
98 & 24 & 4.999 & 0.421 & 2558 & 87552 \\ 
99 & 24 & 4.886 & 0.449 & 2670 & 103767 \\
\midrule
(UC4) \quad
100 & 3 & 3.356 & 0.747 & 2656 & 96695 \\ 
101 & 3 & 3.618 & 0.704 & 2477 & 70112 \\ 
102 & 3 & 3.405 & 0.737 & 2614 & 84743 \\ 
103 & 10 & 4.675 & 0.536 & 2888 & 88983 \\ 
104 & 10 & 4.763 & 0.511 & 2963 & 94337 \\ 
105 & 10 & 4.632 & 0.506 & 2739 & 110944 \\ 
106 & 17 & 4.645 & 0.517 & 2424 & 67638 \\ 
107 & 17 & 4.876 & 0.462 & 2649 & 85094 \\ 
108 & 17 & 4.382 & 0.578 & 2481 & 85904 \\ 
109 & 24 & 4.784 & 0.468 & 2619 & 91658 \\ 
110 & 24 & 4.970 & 0.438 & 2756 & 100208 \\ 
111 & 24 & 4.730 & 0.489 & 2461 & 85470 \\ 
\bottomrule
\end{tabular}
\end{table}

\clearpage
\subsection*{Sample Accession ID}

\begin{table}[bthp]
\centering
\caption{
Sample Accession ID.
}
\begin{tabular}{l|r|l}
Condition & Day & Accession ID\\
\midrule
Control & 0 & DRR157393-157395 \\ 
\quad C: none & 3 & DRR157396-157398 \\ 
\quad N: none & 10 & DRR157399-157401 \\ 
\quad timing: - & 17 & DRR157402-157404 \\ 
& 24 & DRR157405-157407 \\ 
\midrule
AG1 & 3 & DRR157408-157410 \\ 
\quad C: Glucose & 10 & DRR157411-157413 \\ 
\quad N: Ammonium chloride & 17 & DRR157414-157416 \\ 
\quad timing: Abrupt & 24 & DRR157417-157419 \\ 
\midrule
AG4 & 3 & DRR157420-157422 \\ 
\quad C: Glucose & 10 & DRR157423-157425 \\ 
\quad N: Ammonium chloride & 17 & DRR157426-157428 \\ 
\quad timing: Weekly & 24 & DRR157429-157431 \\ 
\midrule
AC1 & 3 & DRR157432-157434 \\ 
\quad C: Cellobiose& 10 & DRR157435-157437 \\ 
\quad N: Ammonium chloride & 17 & DRR157438-157440 \\ 
\quad timing: Abrupt & 24 & DRR157441-157443 \\ 
\midrule
AC4 & 3 & DRR157444-157446 \\ 
\quad C: Cellobiose & 10 & DRR157447-157449 \\ 
\quad N: Ammonium chloride & 17 & DRR157450-157452 \\ 
\quad timing: Weekly & 24 & DRR157453-157455 \\ 
\midrule
UG1 & 3 & DRR169428-169430 \\ 
\quad C: Glucose & 10 & DRR169431-169433 \\ 
\quad N: Urea & 17 & DRR169434-169436 \\ 
\quad timing: Abrupt & 24 & DRR169437-169439 \\ 
\midrule
UG4 & 3 & DRR169440-169442 \\ 
\quad C: Glucose & 10 & DRR169443-169445 \\ 
\quad N: Urea & 17 & DRR169446-169448 \\ 
\quad timing: Weekly & 24 & DRR169449-169451 \\ 
\midrule
UC1 & 3 & DRR169452-169454 \\ 
\quad C: Cellobiose & 10 & DRR169455-169457 \\ 
\quad N: Urea & 17 & DRR169458-169460 \\ 
\quad timing: Abrupt & 24 & DRR169461-169463 \\ 
\midrule
UC4 & 3 & DRR169464-169466 \\ 
\quad C: Cellobiose & 10 & DRR169467-169469 \\ 
\quad N: Urea & 17 & DRR169470-169472 \\ 
\quad timing: Weekly & 24 & DRR169473-169475 \\ 
\bottomrule
\end{tabular}
\end{table}

\end{document}